\documentstyle[epsfig,preprint,tighten,prc,aps]{revtex}

\input{prepictex} \input{pictex} \input{postpictex}
\begin{document}
\draft

\title{Incoherent Photoproduction of $\eta$-mesons from the Deuteron
near Threshold}

\author{A. Sibirtsev$^a$, Ch. Elster$^{a,b}$,
J. Haidenbauer$^a$, and J. Speth$^a$}

\address{
$^a$Institut f\"ur Kernphysik, Forschungszentrum J\"ulich, 
D-52425 J\"ulich \\ 
$^b$Institute of Nuclear and Particle Physics,
Ohio University, Athens, OH 45701 
}

\maketitle

\begin{abstract}
Incoherent photoproduction of the $\eta$-meson on the deuteron is
studied for photon energies from threshold to 800~MeV.  The dominant
contribution, the $\gamma$N-$\eta$N amplitude, is described within an
isobar model. The final state interaction derived from the CD-Bonn
potential is included and found to be important for the description of
the production cross section close to threshold.
Possible effects from the $\eta N$ final state interaction are discussed.
\vspace{1pc}
\end{abstract}
\maketitle

\vspace*{-10.8cm}
\hspace*{11.7cm}
FZJ-IKP-TH-2001-08
\vspace*{10.cm}
%\PACS 13.60.Le \sep 13.75.Cs \sep 14.40.AQ  \sep 14.20.Gk \sep 21.45.+v 

\section{Introduction}
Recent measurements by the TAPS collaboration \cite{Krusche1,Metag,Hejny} at
the MAMI accelerator of the $\eta$-meson photoproduction
on deuterium and helium indicate an enhancement of the total
inclusive cross section at photon energies close to the reaction threshold.
The data were specifically  collected with high statistical accuracy in 
order to clarify the first observation~\cite{Krusche1} of the
rather large total cross section in that energy regime.
It was suggested~\cite{Metag} that such a threshold enhancement 
could result either from the formation of the quasi bound $\eta$-nucleus state
or from the interaction between the final nucleons.
  
Indeed, a strong influence of the final state interaction (FSI) on the cross 
sections of $\pi$, $\eta$, $\eta^\prime$ and $\omega$-meson production in 
nucleon-nucleon ($NN$) collisions was observed in experiments at the IUCF, 
COSY and CELSIUS accelerator 
facilities~\cite{Meyer,Calen1,Calen2,Smyrski,Moskal1,Hibou,Moskal3,Machner}.
With the exception of the $\eta$ channel, those experiments producing mesons
in $NN$ collisions can be described almost perfectly by 
theoretical calculations accounting only for the final state
interactions between the nucleons~\cite{Moskal3,Machner}. In case of
$\eta$ production there is evidence that the $\eta N$ FSI could play a role
as well~\cite{Calen1,Calen2}. Therefore, one might expect that the 
TAPS data can be understood in terms of the strong neutron-proton ($np$) FSI
and possibly an $\eta N$ FSI. 

However, recent calculations~\cite{Arenhovel1} which include the $np$ as
well as the ${\eta}N$ final state interactions 
underestimate the cross section for the reaction $\gamma{d}{\to}np{\eta}$ 
at photon
energies close to the threshold. Within a different approach,
three body calculations~\cite{Arenhovel2} of the reaction 
$\gamma{d}{\to}np{\eta}$ performed by the same authors reproduce the 
main features of the experimental data, but again do not explain the 
rather large total cross section near the reaction threshold.
On the other hand, an older 
calculation of the reaction $\gamma{d}{\to}np{\eta}$ 
performed by Ueda~\cite{Ueda}, which considers the formation of a quasi 
bound $\eta{d}$ state, leads to a much too strong enhancement of the
production cross section close to threshold, and is ruled out by 
the TAPS data. Therefore, the explanation of the TAPS data
is still open and needs further investigations.

Here we evaluate the reaction $\gamma{d}{\to}np{\eta}$ within
the impulse approximation. In addition we account for the FSI between the 
neutron and 
proton by employing the most recent CD-Bonn potential~\cite{Machleidt1}. 
In Sect. 2 we specify the elementary 
$\gamma N\rightarrow \eta N$ amplitude which serves as input for
our calculation of the reaction $\gamma{d}{\to}np{\eta}$. 
Specifically, 
we assume that the elementary $\eta$-production proceeds via the
excitation of the $N^*$ $S_{11}(1535)$ resonance. The free
parameters of our model are fixed by a fit to available data for the reaction
$\gamma p\rightarrow \eta p$.
In Sect. 3 we provide some details about the evaluation of the 
reaction amplitude for $\gamma{d}{\to}np{\eta}$ 
and present results for the impulse approximation as well as
for the inclusion of the FSI in the $np$ system.
Possible effects from the $\eta N$ FSI
are discussed in Sect. 4. In addition we provide predictions for the
angular spectrum and for the momentum spectrum of the produced
$\eta$ meson for selected incident photon energies in the vicinity of
the $\eta$-production threshold. 
The section ends with a brief summary of our results. 

\section{The elementary $\gamma{N}{\to}\eta N$ amplitude.}

The dominant contribution to $\eta$-meson photoproduction from a nucleon
is given by the $N^*$ isobar excitation ~\cite{Knochlein,Benmerrouche}. 
We do not consider the nucleon s-channel pole 
term nor $t$-channel vector meson exchanges, since their contributions
were found to be negligible~\cite{Knochlein,Benmerrouche}. 

The square of the invariant collision energy of the reaction 
$\gamma{N}{\to}N\eta$ is defined as
\begin{equation}
s=m_N^2+2m_N \, E_\gamma ,
\label{etos}
\end{equation}
where $m_N$ and $E_\gamma$ are the nucleon mass and the photon energy, 
respectively. The photon momentum $k$ and the $\eta$-meson momentum $q$
in the center of mass system are given by 
\begin{equation}
k=\frac{s-m_N^2}{2\sqrt{s}}, \,\,\,\,
q=\frac{\lambda^{1/2}(s,m_N^2,m_\eta^2)}{2\sqrt{s}},
\end{equation}
where $m_\eta$ stands for the mass of the $\eta$-meson. The  K\"alen 
function is defined as
\begin{equation}
\lambda(x,y,z)=(x-y-z)^2-4yz.
\end{equation}
The resonant contribution is given  by  
helicity amplitudes in the relevant partial waves 
\cite{Walker,Capstick}, namely
\begin{eqnarray}
\nonumber
A_{l\pm} & = & \pm F A_{1/2}^N \ , \\
B_{l\pm}  &= & \pm F \left[\frac{4}{l(l+2)}\right]^{1/2} A_{3/2}^N \ , 
\label{helicities} \\
\nonumber
C_{l\pm} & = & \pm F C_{1/2}^N \ ,
\end{eqnarray}
where the factor $F$ accounts for the resonance decay into the $N\eta$ 
channel. $l$ denotes the orbital angular momentum.
Taking into account the phase space factor and the 
relativistic Breit-Wigner propagator as introduced in 
Ref.~\cite{PDG} one obtains

\begin{equation}
F= \left[\frac{\Gamma_\eta}{\pi(2j+1)} \, \frac{k}{q}\,
\frac{m_N}{\sqrt {s}} \right]^{1/2} \!\! 
\frac{\sqrt{s}}{M_R^2-s-i\sqrt{s} \ \Gamma} \ .
\label{F}
\end{equation}
Here $M_R$ is the resonance mass, and $\Gamma$ and $\Gamma_\eta$ are the 
total and $R{\to}N\eta$ partial resonance widths, respectively, while
$j$ is the spin of the resonance. 

The standard relation between the Breit-Wigner helicity amplitudes and
electric, magnetic and longitudinal multipoles are given in
Refs.~\cite{Knochlein,Benmerrouche}. 

Following the analysis of pion photoproduction,
we account for the energy dependence of the hadronic
widths~\cite{Manley} in order to satisfy the  threshold
dependence~\cite{Knochlein,Donnachie} of the multipole amplitudes 
of the outgoing meson momentum $q_\xi$. The energy dependence of the
partial width for each final meson $\xi$ is given as
\begin{equation}
\Gamma_\xi = \Gamma_\xi (M_R) \frac{\rho_\xi(\sqrt{s})}{\rho_\xi(M_R)},
\end{equation}
where $\Gamma_\xi (M_R)$ is the $R{\to}N\xi$  partial resonance width
at the resonance pole, while $\rho_\xi$ is given by~\cite{Manley}
\begin{equation}
\rho_\xi(\sqrt{s}){=}\frac{q_\xi}{\sqrt{s}} B_l^2 (q_\xi R), \,\,\,\,
q_\xi{=}\frac{\lambda^{1/2}(s,m_N^2,m_\xi^2)}{2\sqrt{s}} \ .
\label{rhob}
\end{equation}
Here $B_l$ is the Blatt-Weisskopf function for the orbital angular momentum $l$.
The interaction radius was taken as $R$ = 1~fm, and $m_\xi$ stands 
for the mass of the  meson. 
The function $\rho_\xi(M_R)$ in Eq.~\ref{rhob} is 
evaluated at the resonance pole
$\sqrt{s}{=}M_R$. In addition, the total energy-dependent resonance width
is given by the sum over the partial widths of all available final 
states. 

In principle, one may consider the contributions from the resonances 
$P_{11}(1440)$, $D_{13}(1520)$, $S_{11}(1535)$, $S_{11}(1650)$, 
$D_{15}(1675)$, and higher mass resonances to the
photoproduction of $\eta$-mesons \cite{Knochlein} and evaluate the
resonance parameters from the available  differential cross 
section data  and recoil nucleon polarization data \cite{Benmerrouche}.
Contributions from  $S$-wave resonances provide an isotropic
angular spectrum $d\sigma/d\cos\theta$ of $\eta$-mesons, with $\theta$ denoting
the $\eta$-meson emission angle in the c.m. system.
The $P$-wave resonances contribute proportionally to $\cos\theta$, while 
the $D$-wave resonances result in a $\cos^2\theta$ dependence. Although 
contributions from higher partial waves to the total 
photoproduction cross section of $\eta$-mesons can  be very small,
they can be evaluated from the differential $d\sigma/d\cos\theta$
cross section with the help of  interference terms. However, most recent 
data~\cite{Krusche1} for differential cross sections of the 
reaction $\gamma{p}{\to}p\eta$ 
at photon energies from 716 to 788~MeV indicate that, within the experimental 
errors, the angular spectrum is dominated almost entirely by the
$S$-wave distribution. Estimated contributions from
$P$ and $D$-wave resonances can be given only at very low confidence 
level~\cite{PDG3}. Furthermore, data on the nucleon recoil 
polarization, which in principle must be sensitive to the
resonant contribution~\cite{Benmerrouche}, have large uncertainties 
and are thus not significant.

Since there is no strong experimental evidence~\cite{PDG3} 
for contributions to the
$\eta$-meson photoproduction from resonances other than the
$S_{11}(1535)$ resonance in the near-threshold region,
we will consider in the following only this resonance. 
The partial decay widths, $S_{11}(1535){\to}N\eta$ and $S_{11}(1535){\to}N\pi$,
are related to the relevant coupling constant
$g_{RN\xi}$, $\xi{=}\eta,\pi$, by 
\begin{equation}
\Gamma_\xi = \frac{g^2_{RN\xi}}{4\pi} \frac{q_\xi (E_N+m_N)}{M_R}.
\end{equation}
Here the momentum $q_\xi$ and the nucleon energy $E_N$ are evaluated in 
the rest frame of the resonance at the pole position of $S_{11}(1535)$.

Considering only the contribution of the $S_{11}(1535)$ resonance, 
the data for $\eta$-meson photoproduction off protons
can be well fitted with the following resonance parameters
at the $S_{11}(1535)$ pole:
\begin{eqnarray}
M_R=1544~MeV, \,\,\, \Gamma=203~MeV,  \nonumber \\ 
\Gamma_\eta/\Gamma=0.45 \,\,\ , \Gamma_\pi/\Gamma=0.45 \,\,\ ,
\Gamma_{\pi\pi}/\Gamma=0.1 \ . 
\end{eqnarray}
For the electromagnetic helicity amplitudes in Eq.~\ref{helicities} 
we use the values $A^p_{1/2}$ = 0.124 $GeV^{-1/2}$ and 
$A^n_{1/2}$ = -0.1 $GeV^{-1/2}$. 
The result of this fit for the total cross section for 
the reaction  $\gamma{p}{\to}p\eta$ 
is displayed in Fig.~\ref{fi1}. 

\section{The Reaction $\gamma d \rightarrow \eta n p$}

Using the impulse approximation (IA) the amplitude ${\cal M}$ of the
reaction $\gamma{d}{\to}np{\eta}$ for given 
spin $S$ and isospin $T$ of the final nucleons 
can be written as
\begin{equation}
{\cal M}_{IA}{=}A^T(s_1)\phi(\vec{p}_2){-}(-1)^{S+T}A^T(s_2)
\phi (\vec{p}_1),
\end{equation}
where $\phi(p_i)$ stands for the deuteron wave function and 
$p_i$ ($i=1,2$) is the momentum of the proton or neutron in the deuteron 
rest frame. The quantity
$A^T$ denotes the isoscalar or isovector photoproduction 
amplitude at the squared invariant energy $s_N$ given by
\begin{equation}
s_N=s-m_N^2-2(E_\gamma+m_d)E_N+2\vec{k}_\gamma\cdot\vec{p}_N. 
\end{equation}
Our calculation within the framework of the IA is shown
in Fig.~\ref{mainz2} and corresponds to the dashed line. It 
describes the data~\cite{Krusche1} at photon energies above $\simeq$680~MeV 
reasonably well.
Close to the reaction threshold, however, 
the IA result substantially underestimates the data. We take this
as an indication that effects from the ($NN$ and/or $\eta N$)
final state interaction play an important role here. Indeed, 
as already mentioned in the Introduction, it is well known from meson 
production in $NN$ collisions that close to threshold FSI effects lead to 
a significant modification of the cross section.

In meson production in $NN$ collisions 
FSI effects result predominantly from strong $S$-wave interactions 
in the outgoing $NN$ system. Therefore, we will take into account
this contribution for the reaction $\gamma{d}{\to}np{\eta}$. 
The corresponding amplitude is given by 
\begin{equation}
{\cal M}_{FSI}=m_N\int dk k^2 \frac{T(q,k) A^T(s_N) \phi(p_i)}{q^2-k^2+i\epsilon}. 
\end{equation}
Here $q$ is the nucleon momentum in the final $np$ system
and $T(q,k)$ is the half-shell $np$ scattering matrix in the
$^1S_0$ and $^3S_1$ partial waves. 
In the calculations presented here, the half-shell t-matrix is obtained
at corresponding on-shell momenta $q$
from the latest CD-Bonn potential \cite{Machleidt1}, which describes the 
$NN$ data base with a $\chi^2$/datum of about 1.
In order to find out if a high precision
description of the $NN$ data, in our specific case the $NN$ s-waves, 
is  crucial, we carried out the calculations with
an older one-boson-exchange
model, OBEPQ \cite{physrep}, also describing the s-waves reasonably well. 
We found the difference of those two
calculations being negligible.  

The total cross section $\gamma{d}{\to}np{\eta}$ 
including the $np$ FSI in $S$-waves  is displayed in
Fig.~\ref{mainz2} as solid line. Now the model
calculation describes the data~\cite{Krusche1} reasonably well and lies, in
fact, within the experimental uncertainties. As expected, the  FSI
interaction gives rise to a significant increase of the production
cross section close to threshold as is required for getting 
agreement with the data.

\section{Discussion}

In $\eta$-production experiments in $pp$ as well as in $np$ 
collisions one has observed that there is an even stronger enhancement
of the production cross section close to threshold, which cannot
be explained by FSI effects from the $NN$ interaction 
alone~\cite{Calen1,Calen2,Moskal3}. This additional enhancement is, in
general, seen as an indication of FSI effects due to the $\eta N$ 
interaction~\cite{Grishina,Pena}. 
Thus, it may be suggested that similar effects are seen 
in the reaction $\gamma{d}{\to}np{\eta}$. In order to expose
a possible influence from the $\eta N$ FSI we again show the experimental
data in Fig.~\ref{FSIo}, but now divide them by our model calculation,
which includes the enhancement from the FSI between the nucleons. 
Any effects from the $\eta N$ FSI present in the data would then 
reveal themselves as additional enhancement. Indeed, as can be seen in
Fig.~\ref{FSIo}, there is a deviation from our calculation
for energies very close to threshold,
which may be interpreted
as being caused by an $\eta N$ FSI, though the error bars are large.
It is interesting to mention, that the magnitude 
and also the energy range of this deviation are comparable to the effects 
seen in $\eta$-production via hadronic probes. In the reactions 
$pp\rightarrow pp\eta$ as well as in $pn\rightarrow d\eta$ the observed
additional enhancement very close to threshold was a factor of 2 to 3, cf. 
Ref.~\cite{Calen2} and \cite{Calen1}, respectively, and 
the enhancement was limited to excess energies below roughly 15 MeV for the
former and roughly 10 MeV for the latter reaction. In any case, it would
be very useful to have data with higher statistics available at those
energies very close to threshold in order to chart the possible enhancement
due to the $\eta N$ FSI more accurately \cite{Hejny}.  

Angular spectra of $\eta$-mesons
in the photon-deuteron rest frame are shown for different photon energies
in Fig.~\ref{mainz5a}.
The IA calculation underestimates the data at $E_\gamma$=627-665~MeV, 
but already reasonably reproduces experimental results at 665-705~MeV. 

Momentum spectra of the $\eta$-mesons in the $\gamma{-}d$ rest frame 
at different photon energies are displayed in Fig.~\ref{mainz5b}. 
At the lower photon energy, 627-665~MeV, the IA calculation differs 
considerably from the full calculation including FSI. 
The latter leads to a significant enhancement of the yield for larger $\eta$ 
momenta. This is not surprising because in this case the $\eta$-meson 
carries away 
much of the available kinetic energy and the $NN$ system emerges with a 
small relative momentum, and the interaction is particularly strong. 
This enhancement at large $\eta$ momentum is clearly seen in the new still
preliminary data of the TAPS collaboration \cite{Hejny}. 
As the photon energy increases, the difference between the IA and the calculation
including the $NN$ FSI becomes smaller. At a photon energy of 665-705~MeV, 
the effect of the FSI has basically vanished, consistent with the observations 
in Fig.~2.

We would like to emphasize that the theoretical results displayed in 
Figs.~\ref{mainz5a},\ref{mainz5b} represent an average over a finite energy
interval. This is done in order to make the predictions comparable to the
experiments where likewise an averaging over energy bins is 
made \cite{Krusche1,Hejny}. Specifically for the momentum distribution of 
the $\eta$ meson this averaging has a significant influence on the
results. The maximal $\eta$-momentum available at a given fixed 
photon energy for the reaction $\gamma{d}{\to}np{\eta}$ is defined by
\begin{equation}
p_\eta^{max}{=}\frac{\lambda^{1/2}(s,[m_p+m_n]^2,m_\eta^2)}{2\sqrt{s}},
\end{equation}
where $s$ is defined in Eq.~\ref{etos}.
Averaging over the photon energy leads to a smearing of 
$p_\eta^{max}$. Since the $NN$ FSI is most strongly felt for
$\eta$ momenta close to $p_\eta^{max}$ its effect is also smeared
out by averaging over $E_\gamma$ as is the case with the results shown
in Fig.~\ref{mainz5b}. Predictions for a sharp incident photon 
energy show a much stronger structure due to FSI as is exemplified in 
Fig.~\ref{mainz6}. Clearly, this suggests that a high enery resolution in
the experiments is very desirable if one wants to see and study effects
from the FSI. 

\section{Summary}

We calculated the reaction $\gamma d \rightarrow np\eta$ including
the dominant $S_{11}(1535)$ resonance and the neutron-proton final state 
interaction. We find that the impulse approximation reproduces the cross section
for inclusive photoproduction of $\eta$-mesons and the $\eta$-meson angular spectrum
quite well for energies around 680~MeV and higher. At lower energies the consideration
of the FSI between the outgoing nucleons is necessary to describe the relative 
enhancement of the cross-section data with respect to the impulse approximation. 
Though the $NN$ FSI accounts for a large part of the observed enhancement, 
our analysis suggests that there is still a 
remaining discrepancy with regard to 
the data for very small excess energies. This discrepancy is of similar size
as found in the $\eta$-production in $NN$ collisions and may be 
taken as signature of
the $\eta N$ final state interaction very close to threshold. 

\section*{Acknowledgments}
We acknowledge valuable discussions with V. Baru, A. Gasparian, V. Hejny, 
B. Krusche, A. Kudryavtsev, V. Metag, H. Str\"oher and J. Weiss. This work was 
performed in part under the auspices of the U.~S. Department of Energy under contract
No. DE-FG02-93ER40756 with the Ohio University.

\newpage

\begin{figure}
\caption{Total $\gamma{p}{\to}p\eta$ cross section. Experimental data
are from Ref.~\protect\cite{Krusche1}, while the solid line gives our 
result.}
\label{fi1}
\end{figure}

\begin{figure}
\caption{The cross section for inclusive photoproduction of
$\eta$-mesons off deuterium. Experimental data are taken from 
Ref.~\protect\cite{Krusche1}. The dashed line shows the IA
calculation, while the solid line is 
the result with $np$ final state interaction.}
\label{mainz2}
\end{figure}

\begin{figure}
\caption{The cross section for inclusive photon production of $\eta$-mesons 
off the deuteron as a function of the excess energy $\varepsilon$. 
Shown is the experimental cross section divided by the full calculation given 
in Fig.~2.}
\label{FSIo}
\end{figure}

\begin{figure}
\caption{The angular spectra of the $\eta$-meson in the photon-deuteron
rest frame at different the photon energies.  Experimental data are taken from 
Ref.~\protect\cite{Krusche1}. The dashed line shows 
the IA calculation, while the solid line is the result with $np$ FSI.
The theoretical results represent
an average over the given finite energy interval.}
\label{mainz5a}
\end{figure}

\begin{figure}
\caption{The $\eta$-meson momentum spectra in the photon-deuteron
rest frame at different ranges of the photon energies.  
The dashed line shows the IA calculation, while the solid line 
represents the result with $np$ FSI. The theoretical results represent
an average over the given finite energy interval.}
\label{mainz5b}
\end{figure}

\begin{figure}
\caption{The $\eta$-meson momentum spectrum in the photon-deuteron
rest frame at the sharp photon energy of $E_\gamma$ = 660 MeV.  
The dashed line is the result without $NN$ FSI whereas the 
solid line includes it.}
\label{mainz6}
\end{figure}

\newpage 

\begin{figure}
\begin{center}
%\vspace*{-9mm}\hspace*{-4mm}
\psfig{file=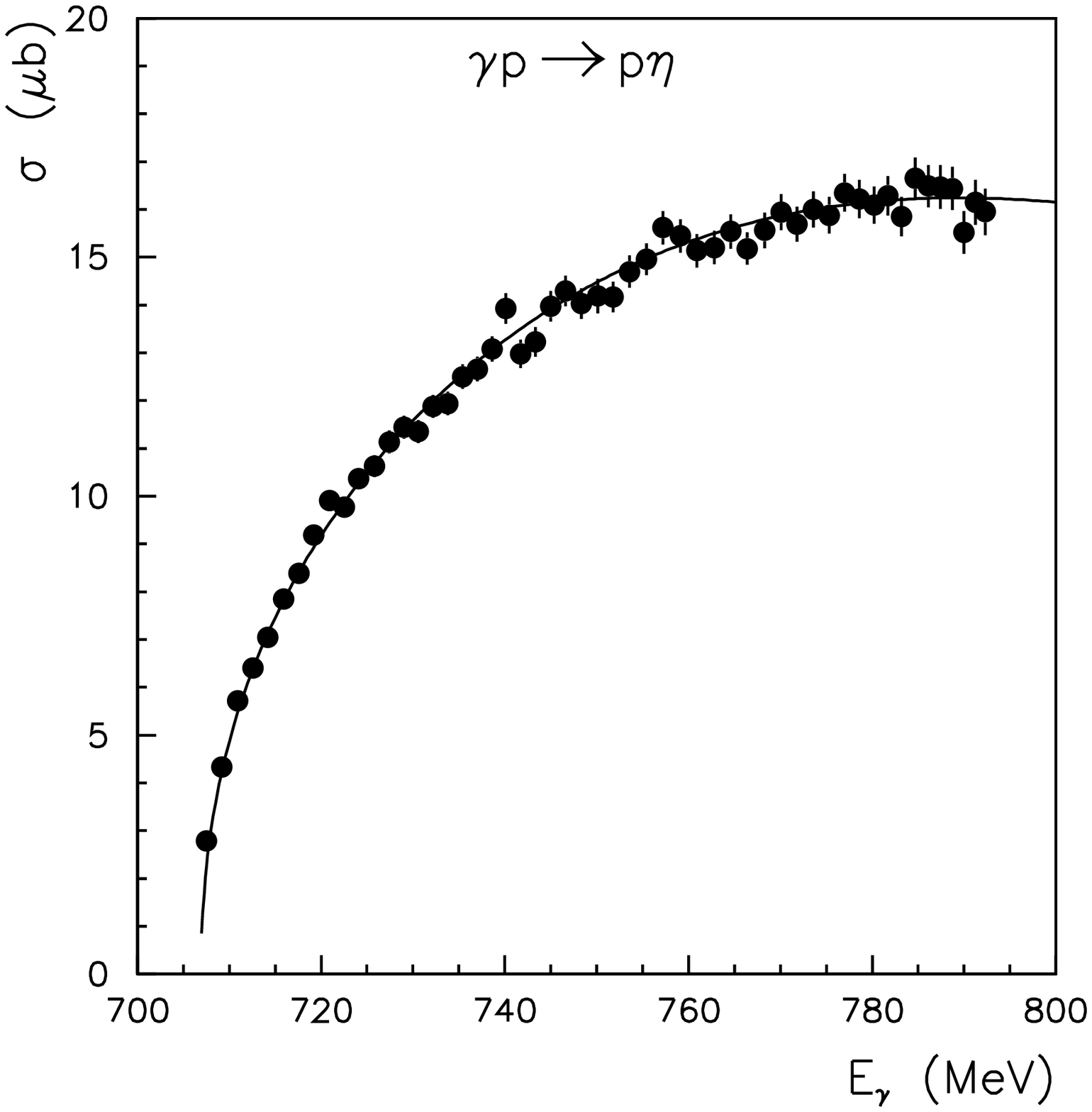,width=11.0cm,height=10.cm}
%\vspace*{10mm}
\center{FIG. 1}
\end{center}
\end{figure}

\begin{figure}
\begin{center}
\psfig{file=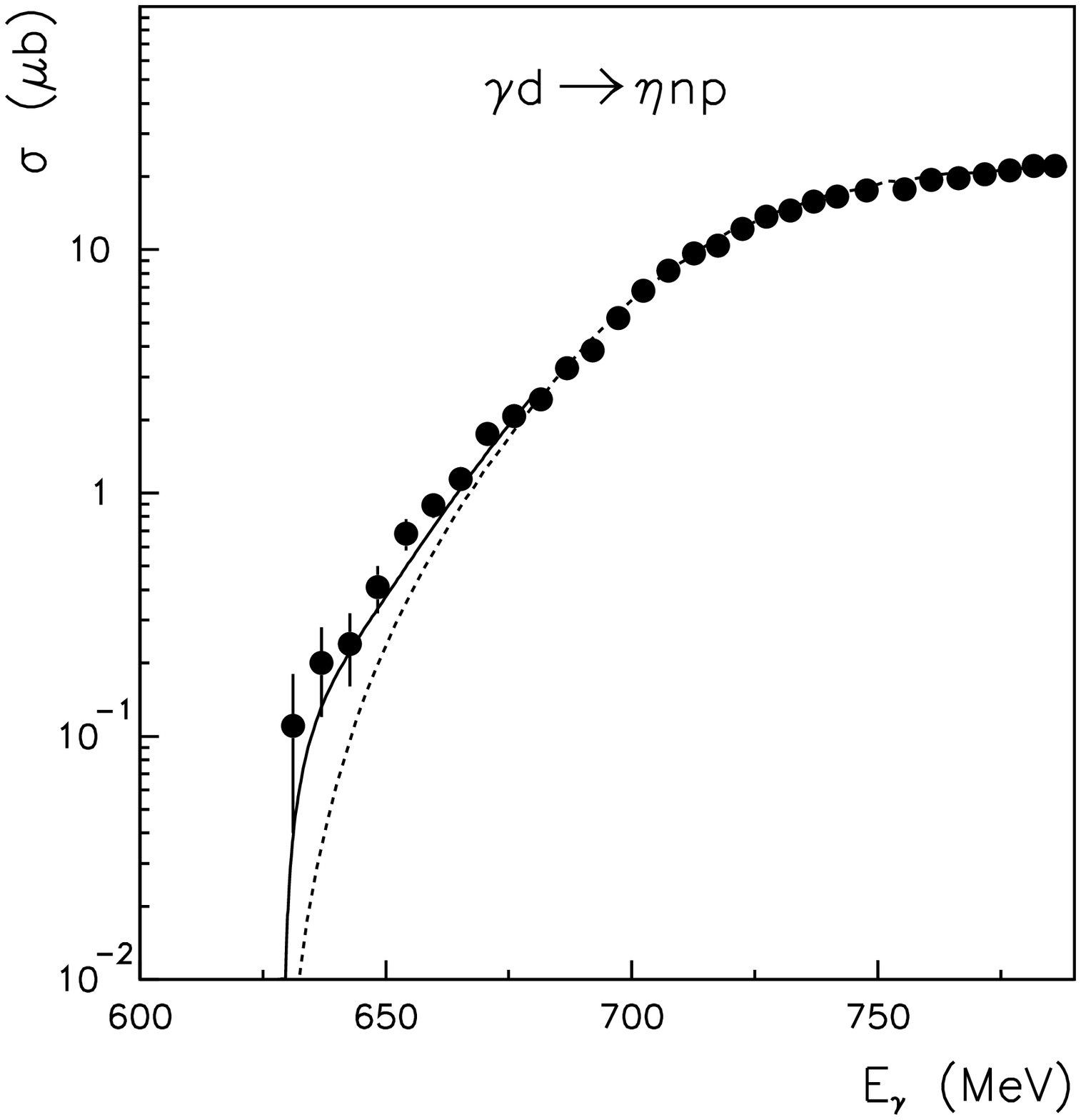,width=11.0cm,height=10.cm}
\center{FIG. 2}
\end{center}
\end{figure}

\newpage

\begin{figure}
\begin{center}
\psfig{file=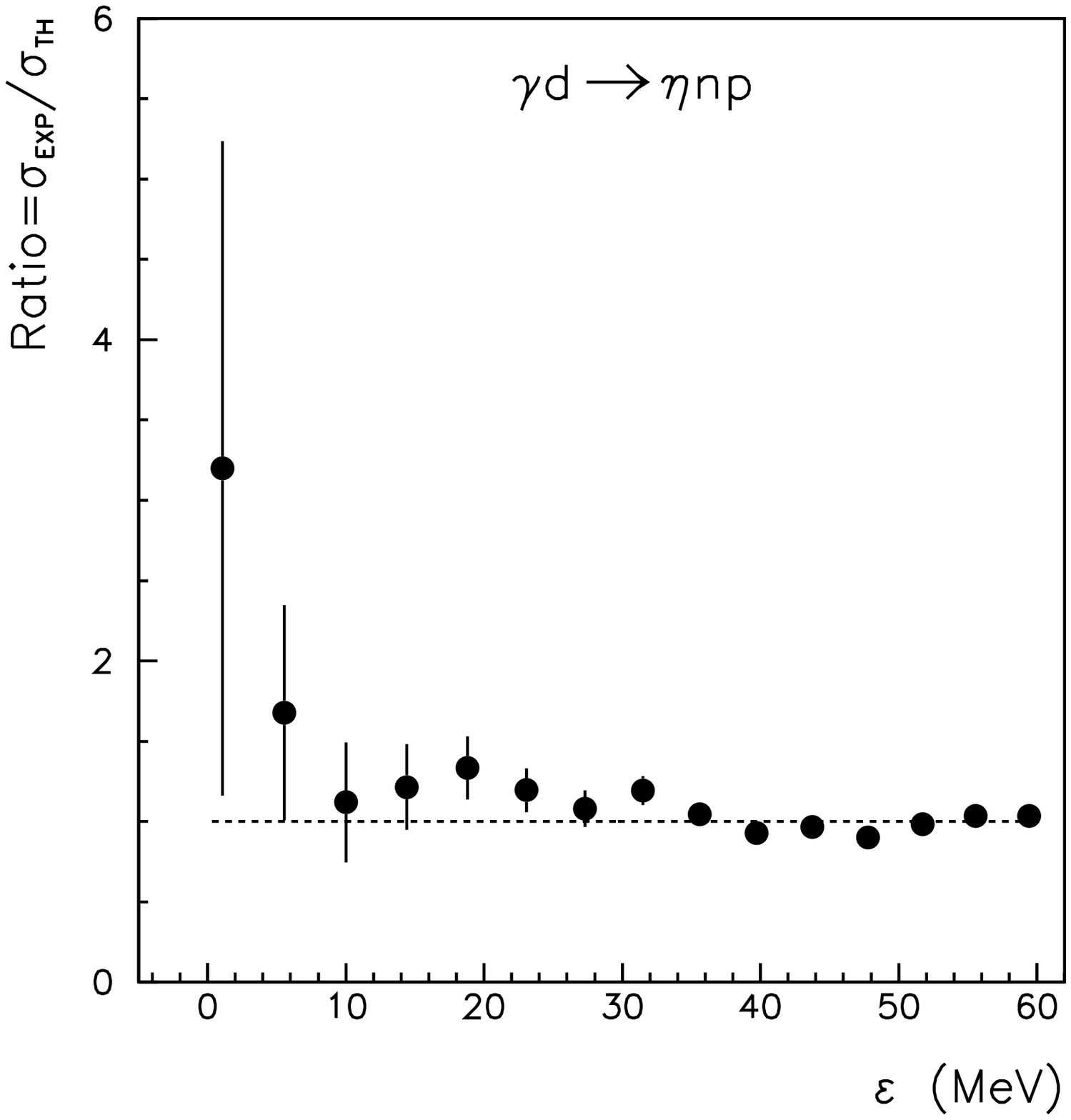,width=11.cm,height=10.cm}
\center{FIG. 3}
\end{center}
\end{figure}

\begin{figure}
\begin{center}
\psfig{file=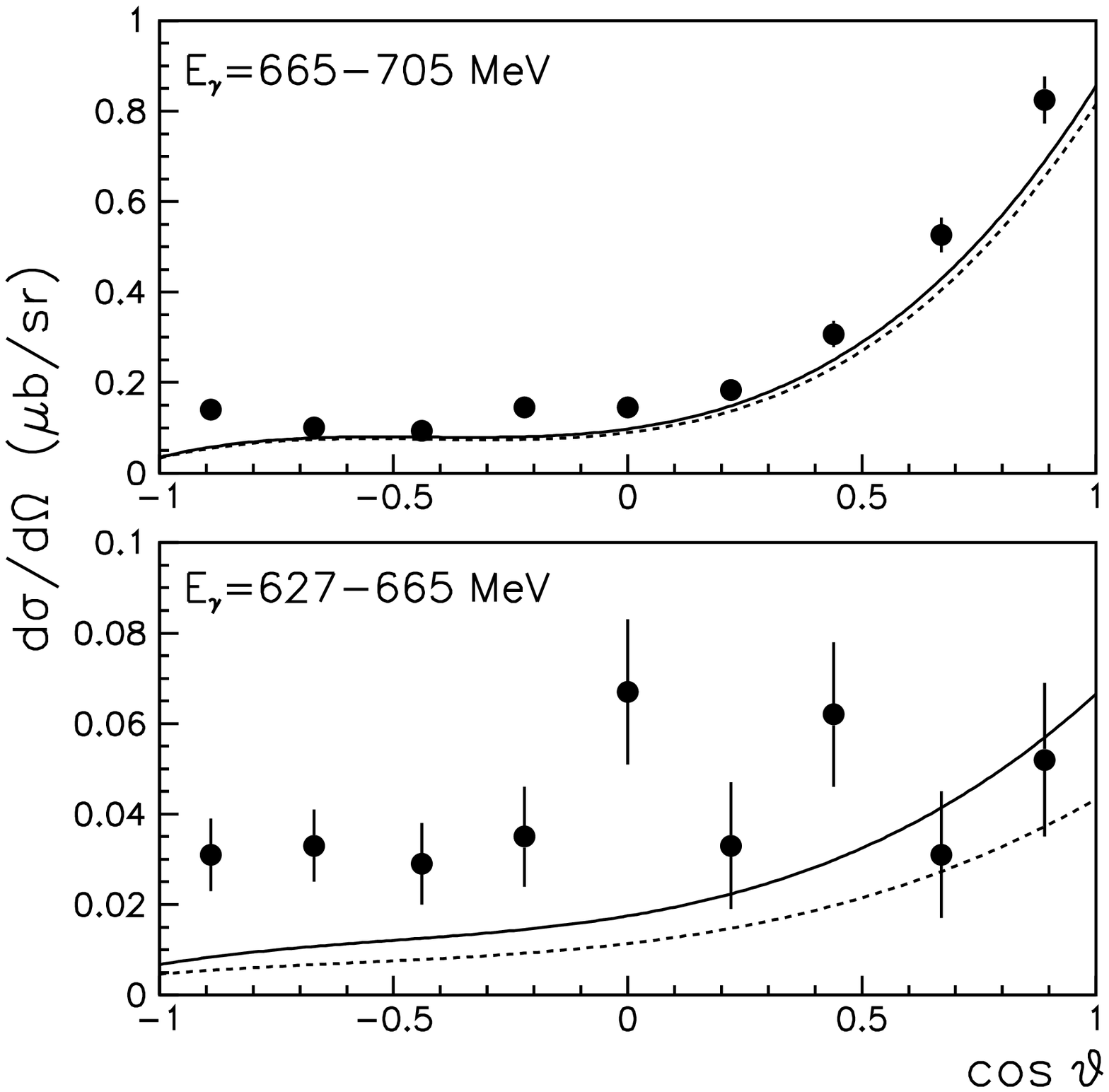,width=11.cm,height=10.cm}
\center{FIG. 4}
\end{center}
\end{figure}

\newpage

\begin{figure}
\begin{center}
\psfig{file=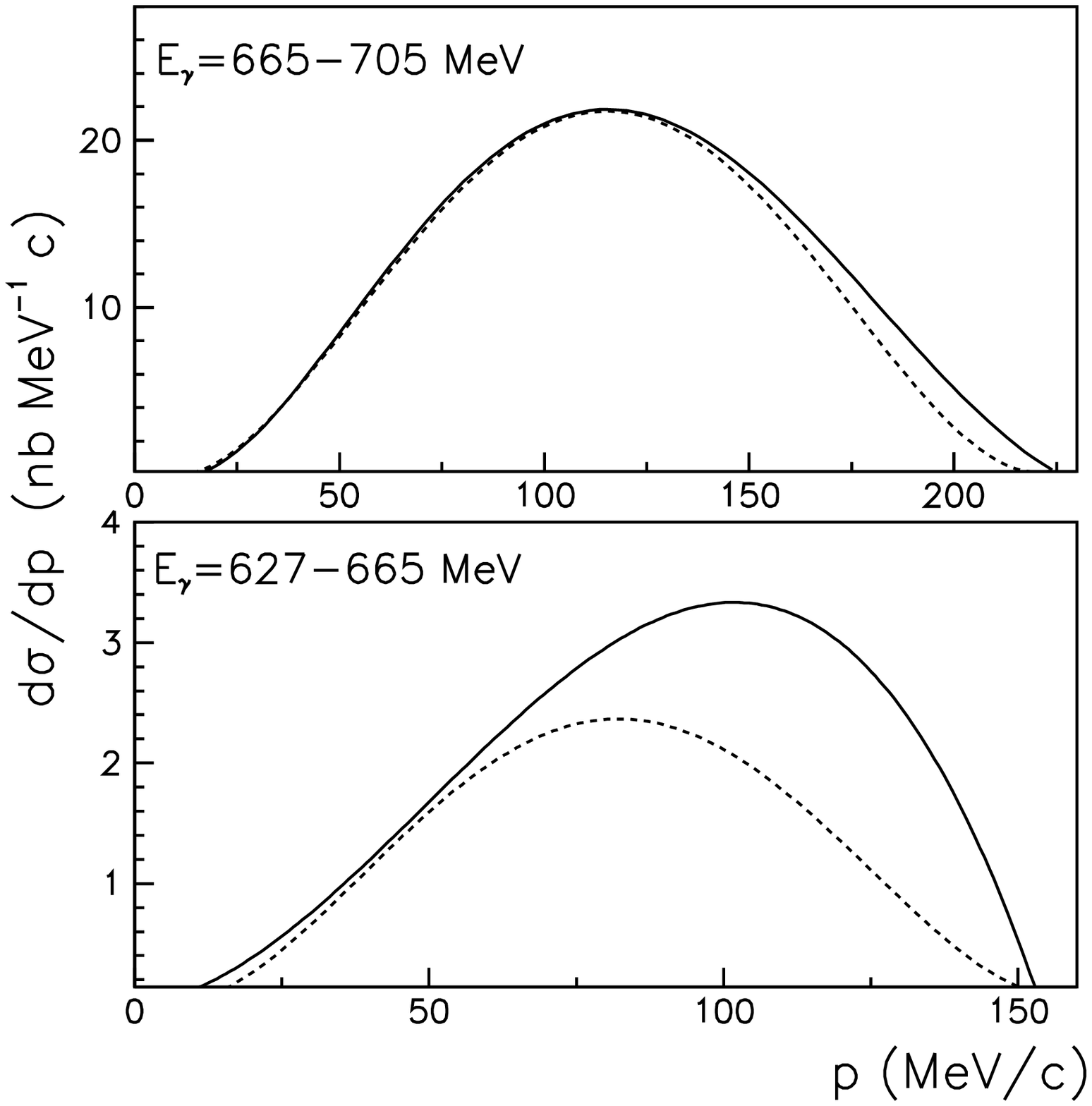,width=11cm,height=10.cm}
\center{FIG. 5}
\end{center}
\end{figure}

\begin{figure}
\begin{center}
\psfig{file=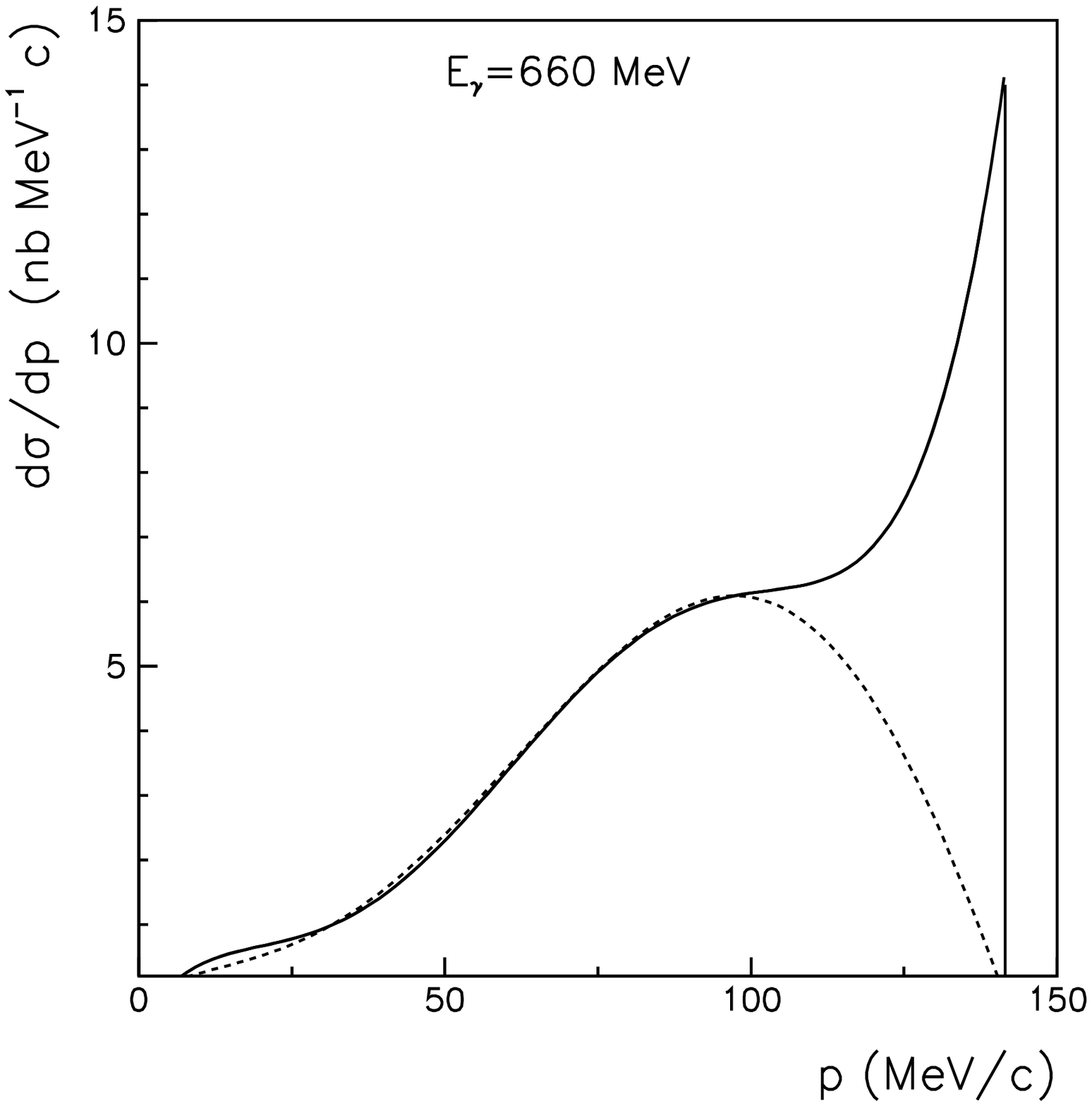,width=11.cm,height=10.cm}
\center{FIG. 6}
\end{center}
\end{figure}


\begin{thebibliography}{00}

\bibitem{Krusche1}
        B. Krusche et al., Phys. Lett. B \textbf{358}, 40 (1995).
\bibitem{Metag}
        V. Metag, in {\it Baryons'98.
Proceedings of the 8th International Conference on the Structure of
Baryons}, edited by D.W. Menze and B.Ch. Metsch
(World Scientific, Singapore 1999), pp. 683.
\bibitem{Hejny}
        V. Hejny et al., in preparation; H. Str\"oher, private communication. 
\bibitem{Meyer}
        H.O. Meyer et al., Phys. Rev. Lett. \textbf{65}, 2846 (1990);
        H.O. Meyer et al., Nucl. Phys. A \textbf{539}, 633 (1992).
\bibitem{Calen1}
        H. Cal\'en et al., Phys. Lett. B \textbf{366}, 39 (1996). 
\bibitem{Calen2}
        H. Cal\'en et al., Phys. Rev. Lett. \textbf{80}, 2069 (1998). 
\bibitem{Smyrski}
        J. Smyrski et al., Phys. Lett. B \textbf{474}, 182 (2000).         
\bibitem{Moskal1}
        P. Moskal et al., Phys. Rev. Lett. \textbf{80}, 3202 (1998); 
        P. Moskal et al., Phys. Lett. B \textbf{474}, 416 (2000). 
\bibitem{Hibou}
        F. Hibou et al., Phys. Rev. Lett. \textbf{83}, 492 (1999). 
\bibitem{Moskal3}
        P. Moskal et al., Phys. Lett. B \textbf{482}, 356 (2000). 
\bibitem{Machner}
        For an overview and further references see, e.g., 
        H. Machner and J. Haidenbauer, J. Phys. G \textbf{25}, R231 (1999).
\bibitem{Arenhovel1}
        A. Fix and H. Arenh\"ovel, Z. Phys. A \textbf{359}, 427 (1997).
\bibitem{Arenhovel2}
        A. Fix and H. Arenh\"ovel, Phys. Lett. B \textbf{492}, 32 (2000). 
\bibitem{Ueda}
        T. Ueda, Phys. Lett. B \textbf{291}, 228 (1992). 
\bibitem{Machleidt1}
        R. Machleidt, Phys. Rev. C \textbf{63}, 024001 (2001).
\bibitem{physrep} 
        R. Machleidt, K.~Holinde, Ch.~Elster,
        Physics Reports {\bf 149}, 1 (1987).
\bibitem{Krusche2}
        B. Krusche et al., Phys. Rev. Lett. \textbf{74}, 3736 (1995).
\bibitem{Dilg}
        W. Dilg, Phys. Rev. C \textbf{3}, 103 (1975).
\bibitem{Klarsfeld}
        S. Klarsfeld, J. Martorell and D.W.L. Sprung, J. Phys.
        G \textbf{10},  165 (1984).
\bibitem{Knochlein}
        G. Kn\"ochlein, D. Drechsel and L. Tiator, Z. Phys. A
        \textbf{352}, 327 (1995)
\bibitem{Benmerrouche}
        M. Benmerrouche and N.C. Mukhopadhyay, Phys. Rev. D \textbf{51},
        3237 (1995).
\bibitem{Walker}
        R.L. Walker, Phys. Rev. \textbf{182}, 1729 (1969).
\bibitem{Capstick}
        S. Capstick and B.D. Keister, Phys. Rev. D \textbf{51}, 3598 (1995).
\bibitem{PDG}
        Particle Data Group, Rev. Mod. Phys. \textbf{48}, 157 (1976).
\bibitem{Manley}
        D.M. Manley et al., Phys. Rev. D \textbf{30}, 904 (1984); 
        D.M. Manley and E.M. Saleski, Phys. Rev. D \textbf{45}, 4002 (1992).
\bibitem{Donnachie}
        A. Donnachie, High Energy Physics, \textbf{V}, Ed. H.S. Burhop,
        New York Academic Press (1972).
\bibitem{PDG3}
        Particle Data Group, Eur. Phys. J. C \textbf{15}, 1 (2000). 
\bibitem{Grishina} 
        V.Yu. Grishina, L.A. Kondratyuk, M. B\"uscher, J. Haidenbauer, C. Hanhart,
        and J. Speth, Phys. Lett. B \textbf{475}, 9 (2000). 
\bibitem{Pena}
        H. Garcilazo and M.T. Pe\~na, Phys. Rev. C \textbf{61}, 064010 (2000).
\end{thebibliography}
\end{document}